# Automated Place Preference Paradigm for Optogenetic Stimulation of the Pedunculopontine Nucleus Reveals Motor Arrest-Linked Preference Behavior

**Authors:** Guanghui Li[1†*], Xingfei Hou[1*], Zhenxiang Zhao[2†]

**Affiliations:**

[1]Department of Neuroscience, Faculty of Health and Medical Science, University of Copenhagen, DK-2200, Copenhagen N, Denmark

[2]Department of Neurology, Henan Provincial People's Hospital, People's Hospital of Zhengzhou University, Zhengzhou 450003, Henan, China.

*Authors contributed equally.

†Correspondence should be addressed to:

Guanghui Li, PhD.

Blegdamsvej, 3B, 2200 Copenhagen N, Denmark

E-mail: guanghui.li@sund.ku.dk

Tel: (45)71892821

†Zhenxiang Zhao, PhD.

Email: zhenxiangzhao@sina.cn

## Abstract

Understanding how the brain integrates motor suppression with motivational processes remains a fundamental question in neuroscience. The rostral Pedunculopontine nucleus, a brainstem structure involved in motor control, has been shown to induce transient motor arrest upon optogenetic or electrical stimulation. However, our current understanding of its potential role in linking motor suppression with motivational or reinforcement-related processes is still insufficient. To further explore the effects induced by PPN stimulations and infer the potential mechanism underlying its role involved in both motor and emotional regulation, we developed a a fully automated, low-cost system combining real-time animal tracking with closed-loop optogenetic stimulation, using the OpenMV Cam H7 Plus and embedded neural network models. The system autonomously detects the rat's position and triggers optical stimulation upon entry into a predefined region of interest, enabling unbiased, unsupervised behavioral assays. Optogenetic activation of CaMKIIα-expressing neurons in the rostral PPN reliably induced transient motor arrest. When motor arrest was spatially paired with a defined region of interest, rats developed a robust place preference after limited training. These results suggest that rostral PPN activation can couple motor inhibition with reinforcement-related behavioral circuitry. Together, our work provides both a technical framework for scalable closed-loop neuroscience experiments and preliminary evidence that the rostral PPN may participate in coordinating motor suppression with motivational processes.

## Structured abstract

**Introduction**

Understanding how the brain integrates motor suppression with motivational processes remains a fundamental question in neuroscience. The rostral pedunculopontine nucleus, a brainstem structure involved in motor control, has been shown to induce transient motor arrest upon optogenetic or electrical stimulation. However, its role in linking motor inhibition to motivational or reinforcement-related behaviors is not fully understood.

**Objective**

To investigate whether activation of CaMKIIα-expressing neurons in the rostral PPN can influence both motor arrest and reinforcement behavior, and to develop an accessible, automated system for closed-loop behavioral experiments.

**Methods**

We designed a fully automated, low-cost, closed-loop system integrating real-time animal tracking with optogenetic stimulation, using the OpenMV Cam H7 Plus and embedded neural network models. The system autonomously detected rat position and delivered optogenetic stimulation when animals entered a predefined region of interest in a conditioned place preference paradigm.

**Results**

Optogenetic activation of CaMKIIα-expressing neurons in the rostral PPN reliably induced transient motor arrest. When stimulation was consistently paired with a specific location in a conditioned place preference task. When motor arrest was spatially paired with a defined region of interest, rats developed a robust place preference after limited training. These results suggest that rostral PPN activation can couple motor inhibition with reinforcement-related behavioral circuitry.

**Conclusions**

Together, our work provides both a technical framework for scalable closed-loop neuroscience experiments and preliminary evidence that the rostral PPN may participate in coordinating motor suppression with motivational processes.

## 1. Introduction

The Pedunculopontine nucleus (PPN), a component of the midbrain motor area located within the Mesencephalic Locomotor Region (MLR), plays a pivotal role in motor initiation and regulation, particularly during exploratory behaviors associated with slow gait (Caggiano et al., 2018; Goñi-Erro et al., 2023). Beyond motor control, the PPN is implicated in various consciousness-related processes, including emotion regulation (Chiba & Murata, 1985; Mena-Segovia & Bolam, 2017), reward and addiction (Forster & Blaha, 2003; Hong & Hikosaka, 2014), arousal and sleep (Garcia-Rill, 1991; Kroeger et al., 2017), path integration, visual processing, and attention (Winn, 2006; Dean, Redgrave, & Westby, 1989). The MLR, a functionally defined nucleus, comprises both the PPN and the Cuneiform nucleus (CnF), with the CnF located anatomically adjacent to the PPN [1]. The functional roles of the PPN and CNF in locomotion, as integral components of the MLR, have been extensively debated. Locomotion initiation and modulation in response to MLR stimulation have been observed across multiple species, including rats, mice, and non-human primates (Goetz et al., 2016; Milner & Mogenson, 1988; Roseberry et al., 2016).

Locomotor behavior is governed by parallel brainstem circuits that interact with emotional and motivational systems to optimize adaptive responses. Within this framework, The PPN and CnF represent two functionally distinct nodes: the CnF facilitates fast, escape-oriented locomotion, while the PPN contributes to slower gait regulation and behavioral halting (Caggiano et al., 2018; Goñi-Erro et al., 2023; Roseberry et al., 2016; Josset et al., 2018). These state transitions frequently emerge under conditions of shifting internal drives or environmental contingencies, suggesting that these structures are more than locomotor relays. Recent electrophysiological evidence has revealed that stimulation of the rostral PPN leads to motor arrest and suppression of hippocampal theta oscillations, a rhythm commonly associated with exploration and fear. Notably, this manipulation does not induce the theta signature characteristic of fear-induced freezing, suggesting that rostral PPN-mediated motor inhibition can occur independently of canonical fear circuits, and may instead represent a unique form of emotionally neutral or alternative affective suppression (Kaur et al., 2025). In parallel, a newly described glutamatergic projection from the lateral hypothalamus (LHA) to the caudal PPN has been shown to prioritize safety over competing behavioral drives. This circuit is recruited during risk-avoidance behavior, even under conditions of hunger or reproductive drive, and is selectively activated by internal motivational states as well as learned threat cues （Krauth et al., 2025). These findings strengthen the view of the PPN as a behavioral gatekeeper, integrating survival-relevant signals to suppress or redirect action.

Importantly, the PPN is also anatomically and functionally connected to dopaminergic systems in the ventral tegmental area (VTA) and substantia nigra pars compacta (SNc)—key nodes in reinforcement learning, reward prediction, and motivational valuation (Mena-Segovia & Bolam, 2017; Hong & Hikosaka, 2014; Lavoie & Parent, 1994; Zhang, Mena-Segovia, & Gut, 2024; Futami, Takakusaki, & Kitai, 1995; Huerta-Ocampo et al., 2021; Floresco et al., 2003; Beier et al., 2015). Through these connectivities, the PPN may influence not only whether movement is executed, but how motor suppression is emotionally appraised or reinforced. However, it remains unclear whether rostral PPN-induced motor arrest is emotionally neutral or capable of modulating learning and motivational preference.

To address this, we hereby present a conditional place preference (CPP) setup which can be combined with real-time optogenetic stimulation in freely moving rats. Based on pre-trained machine learning models, this experimental setup can monitor the activity status of rats and classify different objects or events in the field of view, allowing context-sensitive optogenetic stimulation based on behavioral triggers in real time.

CPP is commonly used to quantify the incentive effects of drug abuse and the reactivation of non-drug-associated activities. Incorporating optogenetics enables precise pairing of contextual cues with stimulation, facilitating in-depth study of reward circuitry and reinforcement behavior (Stuber, Britt, & Bonci, 2012; Stefanik & Kalivas, 2013; Cao, Burdakov, & Sarnyai, 2011). Differentiating from the existing CPP setups, our design highlights the exploratory interactive nature between the animal and objects in the environment. The types of objects and events can be customized flexibly, making the experimental setup affordable, simple, flexible, and universal, which can largely expand the scope of animal behavior research.

We designed two behavioral paradigms to couple optogenetic activation of CamKIIa-expressing neurons in the rostral PPN with place preference. After only 2-3 training sessions, all rats exhibited a significant increase in preference for the stimulation-paired region of interest (ROI), indicating that activation of this rostral PPN neuronal population carries reinforcing properties. These findings suggest that motor arrest induced by rostral PPN activation may be inherently rewarding or motivating.

## 2. Materials and methods

### *2.1 Animals*

Wild-type Long Evans adult male rats (aged 12 to 16 weeks) were used. Rats purchased from Charles River Laboratories were pair-housed in cages standardized with rearing conditions and maintained on a 12:12 light cycle. Rats designated for optogenetic stimulation had free access to food and water during a 7-day acclimation period. All animals were housed individually and were given ad libitum food and supplemented with dietary gel to facilitate recovery and minimize tissue damage for 3 days after the surgery. Additional soft bedding and nesting material were provided during recovery to mitigate stress. At the end of the final video recording, rats were euthanised by intraperitoneal injection of Phenobarbitone.

### *2.2 Surgical procedures and implants*

Rats were anesthetized with 4% isoflurane (Attane vet. 1000mg/g isoflurane) and kept warm with a 37°C heating pad to prevent hypothermia. Subcutaneous analgesia, including Carprofen at a dosage of 0.01ml per 100g body weight (Rimadyl, Bovis vet. 50g/ml), Buprenorphine at a dose of 0.05mg/kg, and Lidocaine at a volume of 0.1ml, was administered to alleviate pain during and after the surgery. An incision was made on the scalp to expose the skull while ensuring the removal of any residual blood from surrounding tissues using sterile cotton swabs and saline. Then, a handheld dental drill was used to roughen the surface of the skull and eliminate small capillaries.

A headplate fabricated with 3D printing served as the base of the crown. A copper wire mesh, measuring approximately 5x5cm in size and tailored to fit the headplate, is meticulously cut to construct a scaffold for the crown. Subsequently, a layer of dental cement is applied onto the copper wire mesh to form the scaffold of the crown. Then affixing the crown securely to the skull by employing Cyanolit glue. After fixing the crown base, bilateral holes with a diameter of 1.4 mm (FST 19008-14, 1.4 mm) were meticulously drilled using stereotaxic guidance for subsequent bilateral injections and fiber implantation. 500nl of rAAV2/9-CamKIIa-ChrimsonR-mScarlet-KV2.1 (Chettih & Harvey, 2019), obtained from Addgene (Addgene 124651) at a concentration of 1×1012 vg/mL, was infused into the rat at a rate of 100 nl/min at target coordinates for rostral PPN (-8 AP, ±2 ML, -7.72 DV relative to Bregma and Dura). After injection, the capillary remained stationary at the injection site for ten minutes, then retracted gradually at an ascending speed of 5 mm/min to avoid potential leaking. Subsequently, an optic fiber (RWD black ceramic ferrule Ø1.25, core size: 200µm;

numerical aperture:0.39; length:10mm) was implanted precisely at the same coordinates as used for injection but positioned approximately 0.1-0.2mm above the DV level employed during injection by descending slowly at a speed of 1mm/min while ensuring stability through fixation with Tetric EvoFlow dental cement. The rats were allowed three weeks post-surgery recovery period to facilitate both their recuperation and viral expression. Following a three-week recovery period to allow for viral expression, optogenetic stimulation was applied to evaluate motor arrest and behavioral effects. Stimulation with 100 Hz pulses at 50% duty cycle was delivered using a fiber-coupled LED source with a maximum power output of 13 mW at the fiber tip. The power level was calibrated to the minimum effective intensity required to elicit a behavioral response (motor arrest), and the final intensity at the fiber tip across animals was kept between 10-13 mW, corresponding to an estimated irradiance of approximately 32-42 $mW/mm^2$ (based on a 200 μm core diameter, NA = 0.39). All animals received light stimulation within this range to ensure experimental consistency.

***2.3 Trajectory analysis***

Rat's movement videos acquired through a standalone camera can be analyzed by a variety of motion analysis software. Here we use DeepLabCut software to accurately track the rat's movement trajectory. We extracted images from the control and test group videos, labeled the rat's position using the rat's center of gravity, and used these images as a dataset to train the deep neural network. In the trajectory plots output by DeepLabCut, the ROI range was defined as from 0 to 1000 along the x-axis. By counting the number of coordinate points that fell within this range, we inferred the amount of time that the rats spent within the ROI. Combining these counts with the total number of coordinate points, derive the percentage of exploration time that rats spent within the ROI.

***2.4 FOMO MobileNetV2 Training***

Given the limited computing resources available for our model, we adapted the FOMO MobileNetV2 network in our study, which is specifically designed for object detection on resource-constrained mobile devices. Compared to the MobileNetV2 network (Sandler et al., 2018; Nguyen, 2020), FOMO MobileNetV2 is highly compact, requiring significantly fewer computational and memory resources, making it suitable for real-time single-target recognition, albeit with a slight accuracy compromise (Jongboom, n.d.). In object detection tasks, FOMO MobileNetV2 extends beyond binary classification to provide detailed information on the precise location, size, and number of target objects.

We used Edge Impulse (https://www.edgeimpulse.com), a cloud-based tool for embedded machine learning workflows, to customize the training parameters and optimize the model for deployment on the OpenMV Cam (Li, Komi, Sørensen, & Berg, 2025). Model performance was evaluated using approximately 160 labeled images for training and 40 labeled images for testing. Bounding boxes were employed to annotate target objects, with the following class labels: background, rat, ROI, and STI (stimulation-triggering indicator). Notably, the STI class was labeled as RIR (rat in region) in the training logs for Rat1 and Rat2, and as rat (detection of target) for Rat3 and Rat4, respectively. We selected FOMO MobileNetV2 0.35 for our model architecture due to its relatively large width multiplier, which enhances accuracy compared to FOMO MobileNetV2 0.1. An illustrative example of the object detection training setup is available at: https://studio.edgeimpulse.com/studio/367850. The model training pipeline includes a demonstration of the labeling strategy, annotation accuracy, and object detection framework applied in the broader experiment.

Following training, the model's object detection capabilities and accuracy are assessed using the F1 Score as the evaluation metric and set the F1 Score threshold set at 95%. Accuracy was used as the evaluation metric; the model accuracy surpassing 95% was adopted in our experiment. Then, we meticulously analyze these misclassifications to discern their impact on real-world task performance and integrate the model into the OpenMV library to facilitate its deployment on OpenMV4 Cam H7 Plus. The model is packaged as a .zip file and deployed to the OpenMV4 Cam H7 Plus via USB connection. OpenMV IDE functions as the driver for the OpenMV4 Cam H7 Plus [30], executes a script that seamlessly integrates object recognition with laser generation. The script is adapted from the operational script provided by Edge Impulse, further tailored to meet the requirement of our task design.

### *2.5 Object detection models performance*

Behavioral experiments were conducted on four rats, two of which were housed in a polycarbonate box while the other two navigated a Y maze with a closed channel. Each rat had a separate model trained for their specific environment. To maintain consistency between the source video and real conditions, all environmental factors such as lighting, angle, and camera field of view were standardized during model usage. For the model used in the polycarbonate box setting, extensive testing revealed optimal performance with a white background; thus white was adopted as the model's background color under this condition. Model performance parameters are detailed in the Supplementary Figure.1 and Supplementary Table I.

## 3. Results

### *3.1 Coupling Optogenetic Stimulation with CPP Task Based on Locomotion Detection and Event Classification*

We adopted the OpenMV Cam H7 Plus camera to synchronize monitoring rats' activity status with optogenetic stimulation. This camera features a high-performance STM32H743II ARM Cortex M7 processor, facilitating tasks including image processing, feature extraction, and model reasoning. Especially, it supports deployment of Convolution

Neural Network (CNN) based object detection models, and its I/O pins are connected to the laser generator for guiding optogenetic stimulation, which provides an ideal convey to coupling CPP tasks with optogenetic stimulation.

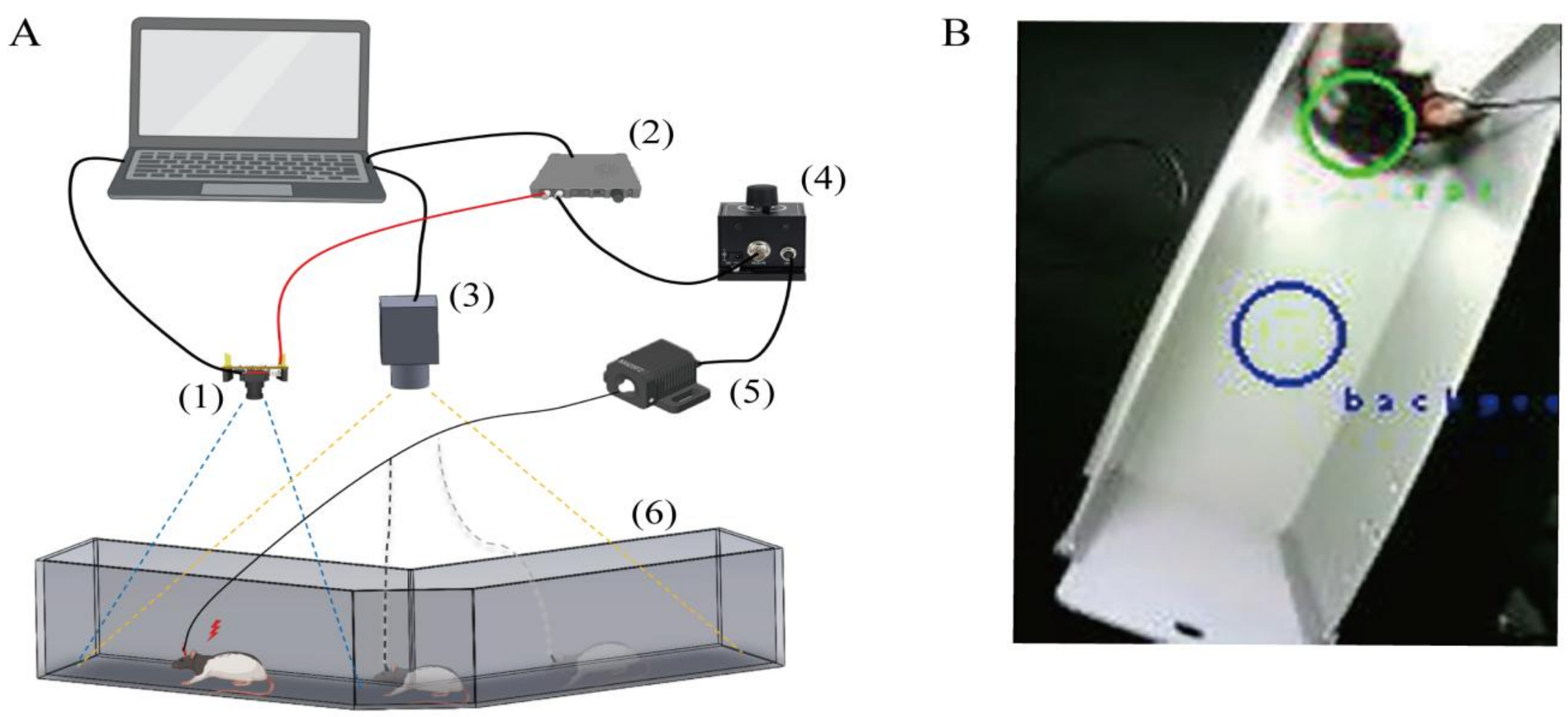

**Figure 1. The setting of coupling optogenetic stimulation with CPP tasks**

(A) Schematic representation of the overall system design. Computer controls OpenMV Cam (1), laser pulse train generator (2) and behavioral recording camera (3). The OpenMV Cam deployed with an object detection model is utilized to supervise the rat's active status, and a trigger signal is sent out to the pulse train generator when the camera detects the target event. The laser pulse train generator, controlled by software named "Pulse Train Generator", enables adjustable laser intensity through a knob (4). The laser (5) is transmitted to the rat's brain by an optical fiber. The behavioral camera is placed above the behavioral chamber to record behavior for further analysis. The polycarbonate behavioral chamber (6) can be substituted with corresponding equipment based on the design of behavioral experiment protocols. (B) A frame from OpenMV Cam. The rat is recognized when it enters one side of the arms in the Y maze and is marked with a green circle, while the background is marked with a blue circle.

As in Figure 1A, the OpenMV Cam camera is placed vertically above the behavioral chamber as a position monitoring device to monitor the animal's movement in real-time, and deployed with a pre-trained CNN-based object recognition and classification model to supervise the animal's active situation (Figure 1B). To deliver temporally precise stimulation upon event detection, the OpenMV Cam was connected to a pulse train generator via a shielding line. Upon recognizing a predefined behavioral event (e.g., the animal entering a Region of Interest), the camera system transmitted a 3.3V trigger signal to the pulse generator (NWT6000 25MHz – 6GHz, TorLabs M625F2, and LEDD1B LED Driver). This configuration allowed real-time activation of optogenetic stimulation to modulate brain activity. Detailed laser stimulation parameters are listed in Supplementary Table II.

***3.2 Adaptation and Operational Features of the Setup in Multiple CPP Tasks***

We designed two behavior tasks with and without visual clues to test the adaptation of the setup. In both tasks, the OpenMV Cam detected the target event and triggered optogenetic stimulation. The behavioral chambers are constructed of polycarbonate, and a part of their compartments is paired with an optogenetics stimulus, referred to as the ROI. The definition of ROI can vary. In one case, a red color block placed underneath the chamber region is functional as a visual clue; the task is designed as the event of the rat shows a preference for the red color block (Figure 2A). In another case, one arm of the Y maze is labeled as ROI without any visual clue, to further exclude possible impairments of visual perception and memory induced by PPN activation (Figure 2D). Once the event is captured by OpenMV Cam and recognized by the built-in pre-trained model, the OpenMV Cam sends out a trigger signal through the 'p.high function', which activated a pulse generator to output a train of laser pulses, thereby initiating stimulation with a 625 nm laser. The laser stimulation consisted of pulsed light at 20 Hz with a 50% duty cycle, delivered for a total duration of 3 second per trigger event. This optical stimulation targeted neurons in the rostral portion of the PPN, inducing a transient global motor arrest. Following each stimulation, the program pauses for 7 seconds to allow the rat to recover and relocate before reinitiating detection and stimulation.

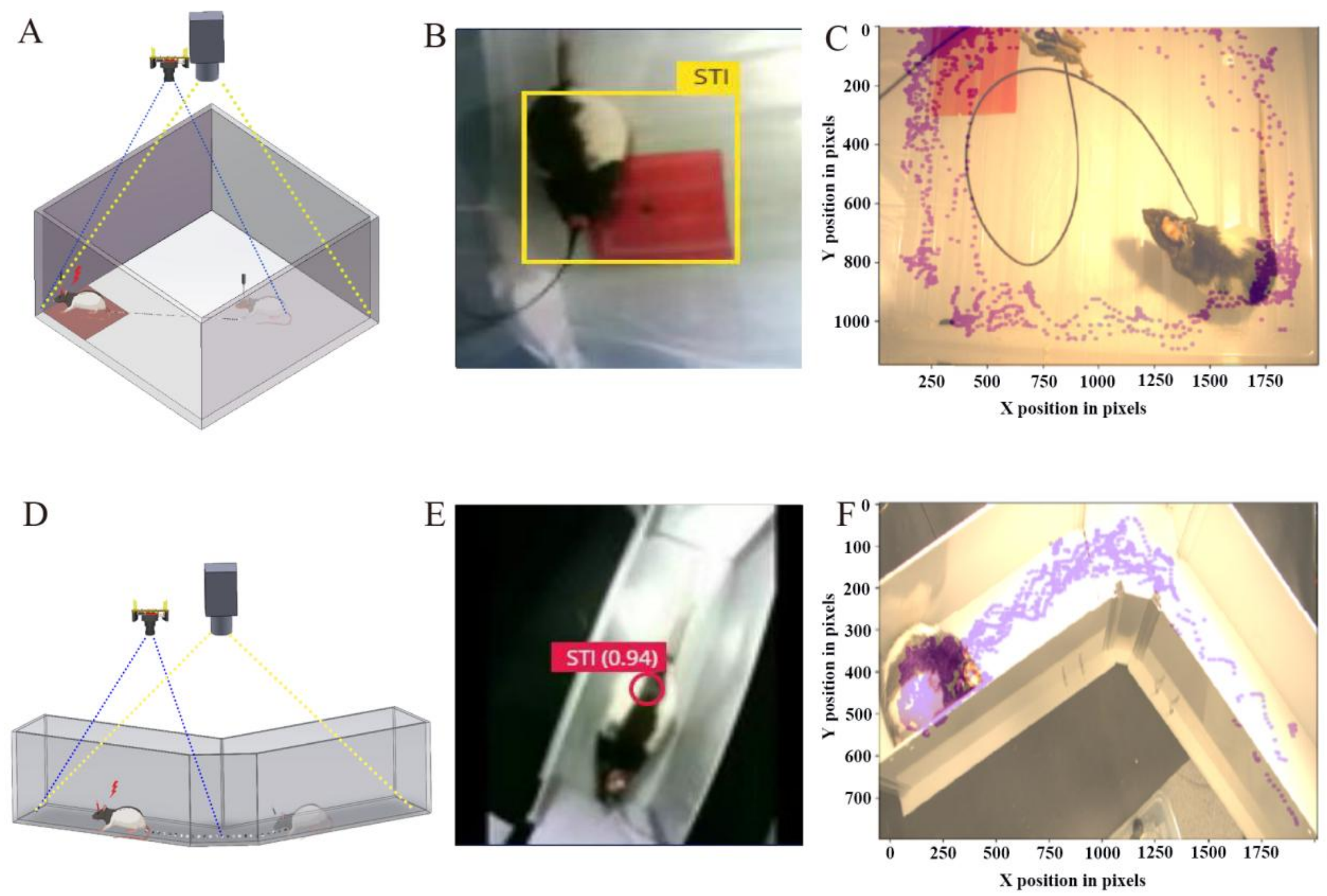

**Figure 2. Establishing the association between stimulus and active status of locomotion in different tasks.**

(A and D) Representation of tasks design. For the task with a visual clue in an open field, a red color block was put underneath the transparent box as a visual clue (A). For the task in bidirectional choice and without a clue, one arm of the Y maze was designated as ROI excluding visual components (D).
(B and E) Labeling of images in Edge Impulse for model training. Events when the rat enters the red color block (B) and one arm of the Y maze (E) are labeled with STI and function as triggering conditions.
(C and F) Trajectories of the animals were recorded from the behavioral camera. Rat in an open field with a visual clue (C), and the bidirectional choice task without a clue (F).

The task for conditional optical stimulation is programmed based on pre-trained machine-learning models to recognize different objects (Figure 3). The CNN model used for object recognition was trained on the Edge Impulse platform using FOMO MobileNetV2 0.35. Training data consisted of frame-extracted videos labeled for object classification (Figure 2B, 2E). Meanwhile, labeling of multiple objects is equally supported and the variety of labeling means that this setup can be applied to different tasks. After training with labeled images, the model was exported and deployed on the OpenMV Cam. Upon event detection, the camera triggered the pulse generator, activating the laser to stimulate the rostral PPN. Movement trajectories were captured by an overhead behavioral camera (Figure 2C, 2F).

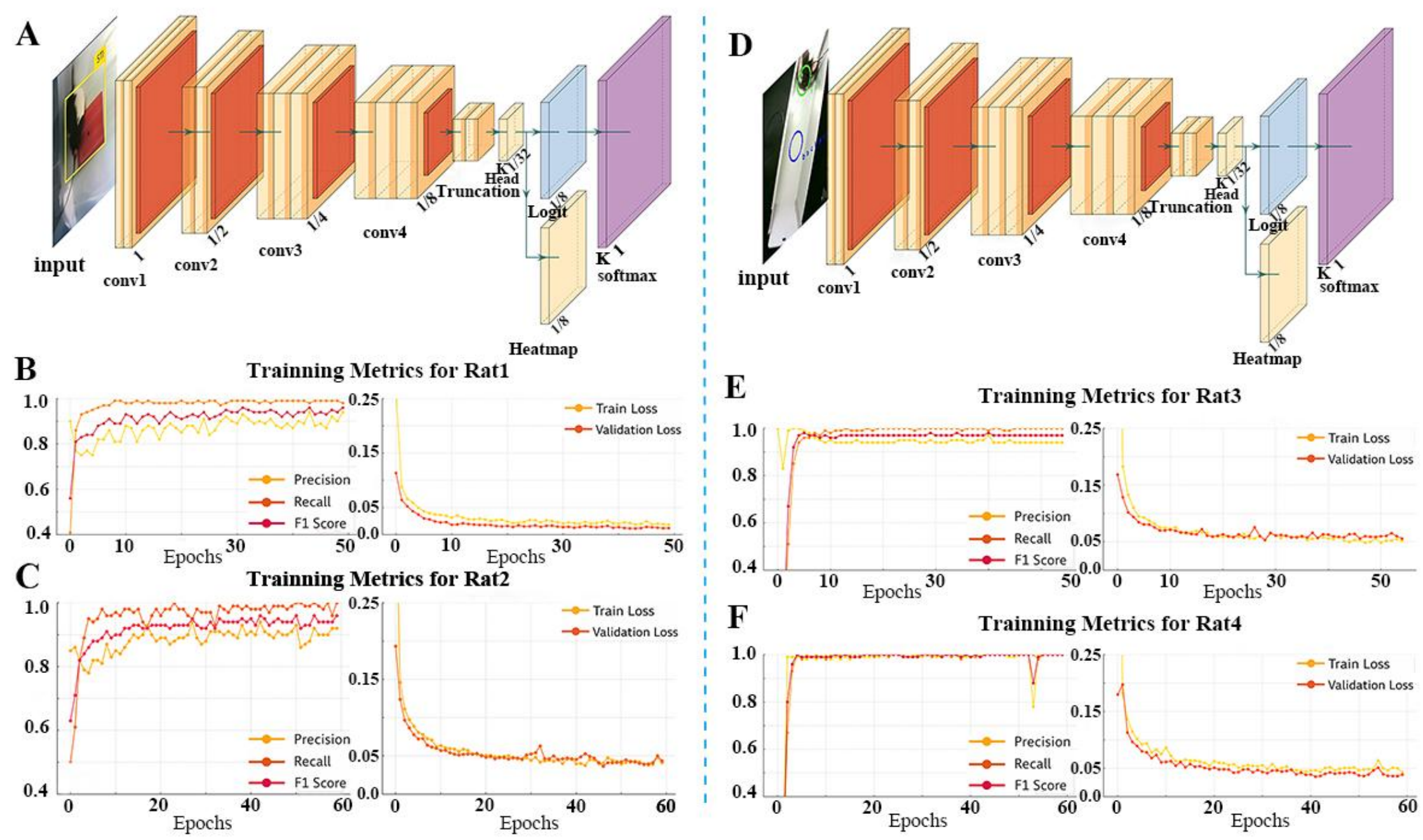

**Figure 3. The architecture of FOMO MobileNetV2 network and the training metrics over epochs.**

(A and D) The architecture of FOMO MobileNetV2 network including the original image as the input layer, MobileNetV2 convolutional blocks (DW Conv1, DW Conv2, DW Conv3, DW Conv4), truncation layer, FOMO-specific layers (head layer, logits layer, softmax layer), and heatmap layer. And the training metrics for Rat1 (B), Rat2 (C), Rat3 (E) and Rat4 (F).

Figures 3A and Figures 3D depict the FOMO MobileNetV2 architecture using PlotNeuralNet [31], illustrating the flow of data from the 160×160×3 input through a series of convolutional and detection-specific layers to a final heatmap output for object localization. The network includes four depthwise separable convolution blocks (DW Conv1 - 4) with increasing channel sizes (32, 64, 128, 256), visually distinguished by custom color coding. FOMO employs a truncated MobileNetV2 backbone, terminating after a designated layer to produce a compact 4×4×n feature map, where n is the number of object classes. This truncation shifts the network's focus from classification to spatially-aware detection. The detection head and logits layers generate raw class predictions, which are then converted to probabilities via a Softmax layer. Unlike conventional object detectors, FOMO outputs a heatmap, enabling efficient object localization without relying on bounding boxes.

Figures 3B and 3C (Rat1 vs. Rat2) and Figures 3E and 3F (Rat3 vs. Rat4) illustrate the training dynamics of the FOMO models over 60 epochs, highlighting efficient learning and generalization across subjects. In all cases, precision, recall, and F1 scores increased rapidly during the early epochs and stabilized above 0.94 from epoch 10 onward, reflecting early convergence and sustained high performance. For both Rat1 (Figures 3B) and Rat2 (Figures 3C), the validation loss decreased steadily alongside the training loss, with minimal divergence between the two models, indicating strong generalization and absence of overfitting. Rat2 achieved slightly higher F1 consistency in the later epochs, suggesting more robust performance in varied instances. Rat3 (Figures 3E) and Rat4 (Figures 3F) also demonstrated rapid loss reduction and plateaued validation curves, but Rat4 exhibited marginally faster convergence, reflected by earlier stabilization of metrics and a more compact loss curve. This may suggest subtle differences in dataset complexity or individual behavior across trials. However, these differences were not statistically significant and are likely attributable to non-biological factors such as scene variability, and lighting conditions or individual differences in fur texture. These variables may influence object detection performance across instances. Nevertheless, the alignment between training and validation losses and the stable performance metrics across all rats underscore the model’s ability to generalize effectively.

### ***3.3 Pedunculopontine Nucleus Induced Motor Arrest Promotes Place Preference Behavior***

PPN is widely involved in the modulation of a variety of cognitive functions, such as arousal, visual input, emotion, and reinforcement learning. It has been shown that PPN activity triggers dopamine release in the nucleus ambiguus and modulates reward signaling critical for reinforcement learning (Forster & Blaha, 2003; Okada et al., 2009; Pan & Hyland, 2005).

To assess whether rostral PPN neuronal activation contributes to the modulation of behavioral reinforcement, rAAV2/9-CamKIIa-ChrimsonR-mScarlet-KV2.1 (Chettih & Harvey, 2019) was injected into the rostral portion of the PPN (Figure 4A). The CamKIIa promoter labels a mixed neuronal population, with a predominance of glutamatergic neurons but additional labeling of cholinergic and GABAergic neurons, as acknowledged in the Discussion. Robust expression of ChrimsonR in the target region is shown in Figures 4B-C, with appropriate rostral targeting and localization confirmed by post hoc histology. Functional targeting of the rostral PPN was verified by optical stimulation, rats were screened for optical responsiveness by assessing motor arrest upon laser stimulation. Only those animals demonstrating reliable, reproducible motor arrest were included in subsequent CPP behavioral experiments (Figure 4D). In a standard paradigm, animals go through three phases: habituation, conditioning of an association between the stimulus and a contextual cue (training), and test phases to evaluate the association. During the habituation (control) phase, rats were placed into the behavioral chamber to explore freely until they showed equal interest in all parts of the chamber. During the Train phase, we activated the OpenMV Cam with a pulse train generator. Whenever the rat entered the ROI, there would be an optogenetic stimulus (20Hz, 25ms pulse width), which resulted in motor arrest and immobilization. The training process was performed for 30 minutes and then the rat was put back into its home cage and rested for 1 hour. The training was repeated 2-3 times per rat. Finally, during the test phase, the OpenMV Cam with pulse train generator was turned off. Rats were allowed to move freely in the behavioral chamber for 30 min to evaluate the training effects. The movement trajectories of the rats at each stage were recorded and used for data analysis.

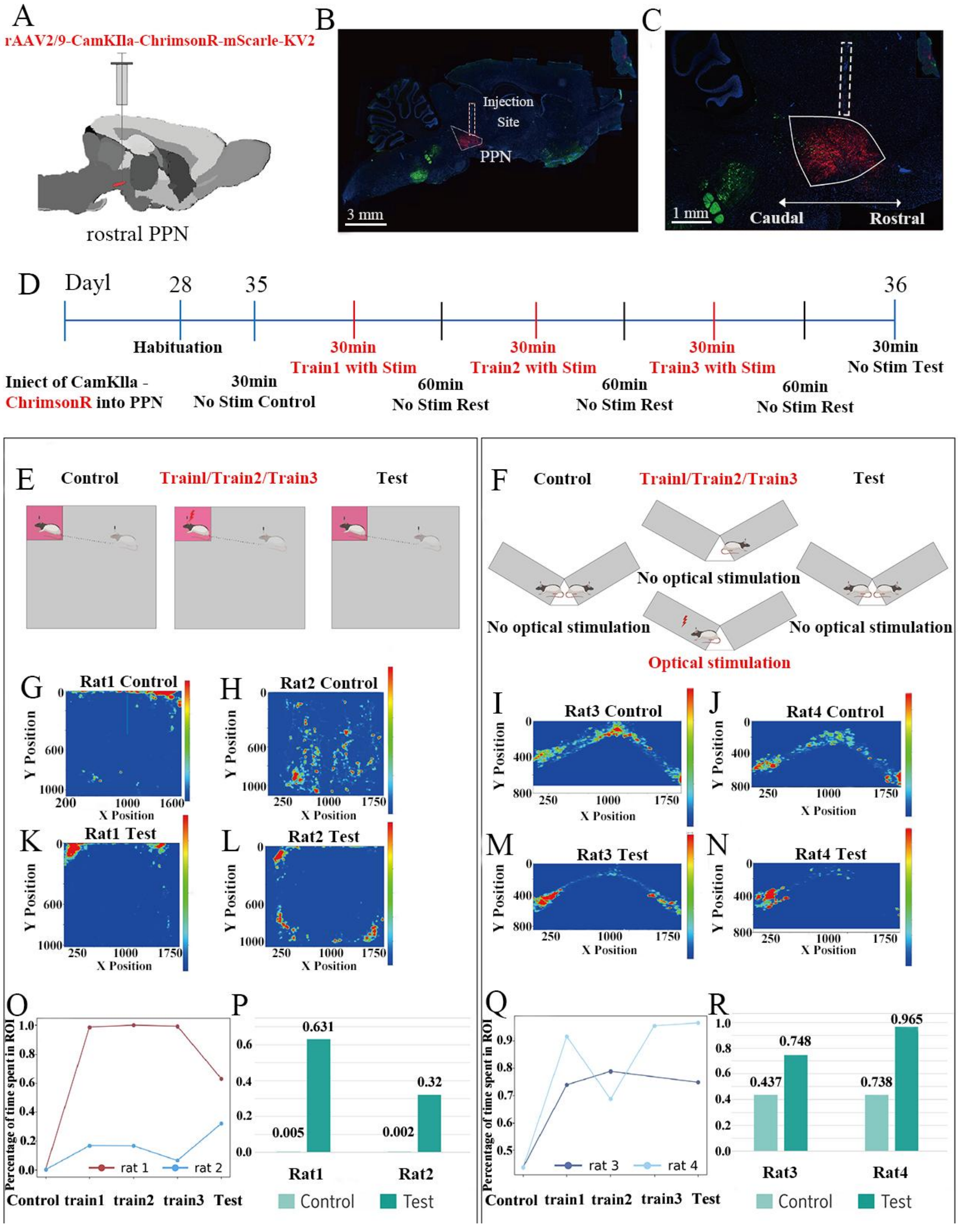

**Figure 4. PPN-induced motor arrest reinforces place preference. (A)** We injected rAAV2/9-CamKIIa-ChrimsonR-mScarlet-KV2.1 into the target region, with histological confirmation of viral targeting showing ChrimsonR expression (red) at the injection site localized to the rostral PPN (B - C), the white dashed line indicates the verified placement of the implanted optical fiber tip relative to the transduced region; (D) CPP Protocol schematic; Behavioral tasks with (E) and without (F) visual cues associated with optogenetic stimulation; rats received 3-second laser pulses (625 nm, 20 Hz, 50% duty cycle) upon entry into ROI; (G_- R) Heatmaps and behavioral quantification generated via DeepLabCut showing spatial occupancy and time spent within the ROI during Control, Training, and Test phases across conditions.

We trained and assessed the positional preferences of rats both with (Figure 4E) and without (Figure 4F) visual cues. During the training phase, rats are repeatedly exposed to stimuli when entering the ROI. The preference/avoidance of rats for ROIs will reflect the reinforcing/aversive effects induced by PPN stimuli.

Motor arrest was induced by optogenetic activation of CamKIIa-expressing neurons in the rostral PPN. In our closed-loop setup, when a rat entered the ROI, a 3-second optogenetic stimulation was delivered, reliably producing a transient motor arrest, followed by a 7-second interval to allow recovery and relocation. Since stimulation was restricted to the ROI, repeated pairing of spatial location and PPN activation led to the formation of a learned association. Videos from the behavior camera were analyzed using DeepLabCut, and heat maps were generated to visualize the percentage of total exploration time spent within the ROI.

Comparisons between control (Figure 4G-J) and test groups (Figure 4K-N) showed that the percentage of time rats spent within the ROIs increased after training with the CPP task, suggesting that the animals developed a preference for the stimulation-paired zone following optogenetic activation of the rostral PPN. This preference development was further supported by trajectory-based analysis (Figure 4O-R), which showed a general upward trend in ROI occupancy across training sessions in most rats. Notably, some animals (e.g., Rat 2) exhibited peak ROI occupancy during the test phase, while others (e.g., Rats 1 and 4) reached maximal values during training. These patterns suggest that optogenetic stimulation of the rostral PPN induces a reinforcing effect, although the expression of preference may vary across individuals and sessions.

These findings suggest that activation of rostral PPN neurons labeled by the CamKIIa promoter is sufficient to drive both motor arrest and reinforcement learning, although the precise contribution of glutamatergic versus non-glutamatergic populations remains to be further defined.

## 4. Discussion

This study was fundamentally motivated by a scientific question: to understand how activating neuronal populations in the rostral PPN affects behavior, specifically motor arrest and reward processing. However, addressing this question required precise, scalable, and minimally intrusive behavioral monitoring and stimulation in a cost efficient manner. Therefore, we developed a low-cost, easy-to-build CPP-optogenetics integration device that enables unsupervised behavioral experiments using a real-time, embedded neural network-based detection model on the OpenMV4 Cam H7 Plus. This system allowing fully autonomous operation without researcher supervision. Objective and automated data collection are essential for ensuring accurate, unbiased analysis of reward and motivational behaviors (von Ziegler, Sturman, & Bohacek, 2021; Brown, 2024; Zhang et al., 2019). Our trained neural network reached high validation (97.1%) and test accuracy (91.1%) in identifying rats during CPP trials. The sysytem's end-to-end latency was measured at 770 ms, including frame capture, inference, and TTL triggering using a quantized int8 model running on a Cortex-M4F (80 MHz). This delay remained consistent during deployment and was sufficient for most ROI-triggered activations within our CPP paradigm. While rapid transient entries (under 1s) could theoretically occur without triggering stimulation, we found that the majority of rat-ROI interactions lasted longer, enabling reliable optogenetic activation. The stability and automated nature of this system enhance the reliability of behavioral experiments, addressing critical concerns about objectivity and reproducibility in research (Crabbe, Wahlsten, & Dudek, 1999; Mandillo et al., 2008). By taking advantage of trained models, the system can identify the rat's position in real-time and respond to the rat's interaction with the ROI quickly and accurately.

In our implementation, the object detection model was trained individually for each rat. This approach minimized variability introduced by individual differences such as fur texture, lighting conditions, and background contrast, thereby maximizing detection accuracy and tracking reliability in real time. Each model was trained using approximately 150-250 manually annotated frames, which were sufficient to achieve high validation accuracy (>95%) and robust performance during closed-loop optogenetic tasks. While individualized training ensured precision in this study, the model framework is scalable. By increasing the dataset to include multiple subjects across varying experimental conditions, it is feasible to train a generalized detection model capable of performing accurately across rats without requiring retraining for each new subject. This possibility supports the broader applicability and scalability of the system in future behavioral neuroscience experiments.

The test of the adaptation in Multiple CPP Tasks indicated that this system possesses many characteristics, such as multi-scene, simple operation, high accuracy, and automation. By utilizing an open-source recording system and free analysis software, users just need to construct their behavior chamber and definite the target tasks. Therefore this setup provides a practical and affordable alternative to commercial solutions, especially when paired with open-source tracking and analysis software. Although the OpenMV Cam has not yet been tested for tracking the movements of multiple rats, nor has it been tested for the recognition of fine animal movements in our system, it still has the possibility to accomplish these tasks functionally. Furthermore, although conventional optogenetic approaches (wired light delivery) were used in our cases, the design applies equally to wireless optogenetics systems such as NeuroLux and CerebraLux, enabling more naturalistic behaviors and enhanced data collection in complex paradigms.

Furthermore, the system enabled novel insights into the function of the PPN in behavioral reinforcement and motor arrest in our study. As part of the dual response systems within the mammalian superior colliculus (SC), the role of PPN in orienting behaviors and threat responses was first discussed in the year of 1988 and was proposed that PPN integrates sensory information to guide both routine, event-driven actions and emergency, reflex-like responses (Dean, Redgrave, & Westby, 1989). Since then, lines of evidence suggested that PPN is a key component of the MLR and is critically involved in regulating motor control and locomotion. Selective stimulation of certain parts or specific subpopulations of the PPN can lead to different movement outputs. Its ability to influence movement depends on the neuronal subpopulations and their connections within the motor and basal ganglia circuitry. Low-frequency stimulation (20 – 25 Hz) of the PPN in humans with Parkinson's disease (PD) has been shown to reduce gait freezing and improve gait and postural control, especially for symptoms like freezing of gait (Thevathasan et al., 2011; Tykocki, Mandat, & Nauman, 2011; Yu et al., 2020). This effect is believed to be due to the activation of PPN cholinergic neurons, which project to motor regions in the brainstem and spinal cord, helping to initiate and maintain locomotion (Thevathasan et al., 2018; Gut & Winn, 2016; Lau et al., 2015; Takakusaki, 2017).

Further study by activating PPN glutamatergic neurons showed regulation of slow locomotion, while the neighboring CnF is responsible for faster gaits (Caggiano et al., 2018), which provides direct experimental evidence that PPN stimulation facilitates locomotion by engaging subpopulations, and their downstream effects on spinal motor circuits. A distinct population of excitatory neurons in the PPN that express the transcription factor Chx10 was identified in PPN, optogenetic activation of Chx10+ neurons in the PPN induces global motor arrest in mice, the motor arrest is transient and showed without any apparent impact on consciousness or the ability to process sensory input, which is hypothesized to be a response to salient environmental cues or a preparatory state for subsequent actions (Goñi-Erro et al., 2023).

Given PPN's multiple roles in consciousness-related functions and movement modulation, especially its involvement in reward processing and movement (Forster & Blaha, 2003; Okada & Kobayashi, 2013), which indicated that activation of PPN excitatory neurons induced motor arrest may also be implicated in reward-related behavior. Investigating the association of reward-related behavior with motor arrest could help to define the functional context of this phenomenon.

In this study, we targeted CamKIIa-expressing neurons in the rostral part of PPN, rats just after 2-3 times training process showed a robust place preference behavior, this finding suggest that activation of this neuronal population not only suppresses locomotion but also engages reinforcement mechanisms, implicating the PPN as a hub coordinating both motor and motivational systems. Glutamatergic PPN neurons are known project to brainstem motor centers, such as the medullary reticular formation, which are crucial for motor execution. Excitation of these pathways may disrupt locomotor activity by overriding motor output from the basal ganglia to the spinal cord, leading to motor arrest. Meanwhile, PPN inputs to the ventral tegmental area (VTA) and substantia nigra pars compacta (SNc) modulate dopamine activity, which underlies reinforcement learning, reward, and motivation valuation. This dual connectivity of the PPN allows it to serve as a "behavioral gatekeeper" coordinating motor and motivational states based on environmental demands or internal drives.

However, consistent with prior findings, our use of ChrimsonR under the CamKIIa promoter was not strictly selective for glutamatergic neurons. In a recent study using the same viral strategy targeting the anterior rostral PPN, approximately 54% of transduced neurons were VGluT2+ (glutamatergic), while 25% were PAX2$^{+}$ (GABAergic), and a small fraction were cholinergic, some of which co-expressed VGluT2 [15]. This distribution reflects a predominant, but not exclusive, glutamatergic transduction profile. Given our similar targeting coordinates and viral strategy, partial recruitment of non-glutamatergic populations is likely. Although the behavioral outcomes align with prior findings in mice, where a subset of PPN glutamatergic neurons expressing the transcription factor Chx10 was shown to induce transient global motor arrest upon optogenetic activation (Goñi-Erro et al., 2023). This limitation should be considered when interpreting cell-type specificity of our study. Future studies using Cre-dependent viral systems in VGluT2-Cre, ChAT-Cre, or GAD2-Cre lines are necessary to delineate the distinct contributions of PPN subtypes to motor and motivational control.

Moreover, our findings of reinforcement-linked motor arrest contrast with recent work, which identified GABAergic PPN neurons that inhibit SNc and VTA dopamine neuron activity, resulting in valence-neutral motor suppression (Zhang, Mena-Segovia, & Gut, 2024). Together, these data support a dual model of PPN function, in which inhibitory neurons gate movement broadly without emotional encoding, while CamKIIa-expressing (predominantly glutamatergic) neurons couple motor inhibition with affective or motivational significance. Disentangling these functionally distinct subpopulations will be key to understanding how the PPN orchestrates behavioral state transitions under different environmental and internal conditions.

Finally, PPN activation offers therapeutic potential across a range of neurological and psychiatric conditions, leveraging its influence on motor control, reward processing, sleep regulation, and cognitive functions. In our study, PPN stimulation showed an effect of strong preference or addiction underscore the need for precise circuit targeting to avoid unintended reward or dependency-related side effects. Future research should focus on refining these interventions, understanding side effects, and tailoring approaches to individual conditions for treating related diseases.

**Funding**

This work was partly supported by the Lundbeck Foundation (Grant No. R366-2021-233), the Start-up Research Grant from Henan Provincial People's Hospital (Grant No. ZC20250077), and the China Scholarship Council (Grant No. 202009110098).

**Declaration of competing interest**

The Authors confirm that there are no conflicts of interest.

**Ethics declaration**

All experimental procedures followed ARRIVE guidelines as well as Council of the European Union regulations (86/609/EEC) and were authorized by the Danish Veterinary and Food Administration (Animal Research Permit No. 2024-15-0201-01739).

**Consent to participate**

Not applicable

**Consent for publication**

All authors have reviewed the final manuscript and consent to its publication.

**Acknowledgements**

The authors would like to express our sincere gratitude to Madelaine Bonfils for her skilled assistance with the animal surgeries and to Professor Rune W. Berg for his support; their contributions to this work are deeply appreciated.

**Data availability**

All source data generated and analyzed in this study are freely and openly accessible without restriction.

The full dataset (~67 GB), comprising all raw video recordings and DeepLabCut-extracted trajectory files from both habituation and test phases for each rat, is hosted at: https://zenodo.org/me/uploads?q=&f=shared_with_me%3Afalse&l=list&p=1&s=10&sort=newest

To further enhance transparency, the complete training logs, model configuration files, and deployment code are publicly accessible at: https://github.com/SapphireV/OpenMV-Cam-Rat-s-Position-Detection.git

**CRediT authorship contribution statement**

**Xingfei Hou**: Data curation, Investigation, Formal analysis, Methodology, Writing-Original Draft, Writing-Review & Editing; **Zhenxiang Zhao**: Conceptualization, Funding acquisition, Validation, Visualization, Writing-Review & Editing; **Guanghui Li**: Conceptualization, Formal analysis, Funding acquisition, Methodology, Project administration, Resources, Software, Supervision, Visualization, Writing-Original Draft, Writing-Review & Editing.

**Supplementary data**

Supplementary material includes 1 figures and 2 tables are available with the online version of this manuscript.

## References

[1] Caggiano, V., Leiras, R., Goñi-Erro, H., Masini, D., Bellardita, C., Bouvier, J., ... Kiehn, O. (2018). Midbrain circuits that set locomotor speed and gait selection. Nature,

553, 455–460. https://doi.org/10.1038/nature25448

[2] Goñi-Erro H, Selvan R, Caggiano V, Leiras R, Kiehn O. Pedunculopontine Chx10(+) neurons control global motor arrest in mice. Nat Neurosci 2023;26:1516–28.Goñi-Erro, H., Selvan, R., Caggiano, V., Leiras, R., & Kiehn, O. (2023). Pedunculopontine Chx10+ neurons control global motor arrest in mice. Nature Neuroscience, 26, 1516–1528. https://doi.org/10.1038/s41593-023-01396-3

[3] Chiba T, Murata Y. (1985). Afferent and efferent connections of the medial preoptic area in the rat: a WGA-HRP study. Brain Res Bull, 14, 261–272. https://doi: 10.1016/0361-9230(85)90091-7

[4] Mena-Segovia, J., & Bolam, J. P. (2017). Rethinking the pedunculopontine nucleus: From cellular organization to function. Neuron, 94, 7–18. https://doi.org/10.1016/j.neuron.2017.02.027

[5] Forster, G. L., & Blaha, C. D. (2003). Pedunculopontine tegmental stimulation evokes striatal dopamine efflux by activation of acetylcholine and glutamate receptors in the midbrain and pons of the rat. European Journal of Neuroscience, 17, 751–762. https://doi.org/10.1046/j.1460-9568.2003.02511.x

[6] Hong, S., & Hikosaka, O. (2014). Pedunculopontine tegmental nucleus neurons provide reward, sensorimotor, and alerting signals to midbrain dopamine neurons. Neuroscience, 282, 139–155. https://doi.org/10.1016/j.neuroscience.2014.07.002

[7] Garcia-Rill, E. (1991). The pedunculopontine nucleus. Progress in Neurobiology, 36, 363–389. https://doi.org/10.1016/0301-0082(91)90016-T

[8] Kroeger, D., Ferrari, L. L., Petit, G., Mahoney, C. E., Fuller, P. M., Arrigoni, E., ... Scammell, T. E. (2017). Cholinergic, glutamatergic, and GABAergic neurons of the pedunculopontine tegmental nucleus have distinct effects on sleep/wake behavior in mice. Journal of Neuroscience, 37, 1352‒1366. https://doi.org/10.1523/JNEUROSCI.1405-16.2016

[9] Winn P. (2006). How best to consider the structure and function of the pedunculopontine tegmental nucleus: evidence from animal studies. Journal of the neurological sciences, 248(1-2), 234‒250. https://doi.org/10.1016/j.jns.2006.05.036

[10] Dean, P., Redgrave, P., & Westby, G. W. (1989). Event or emergency? Two response systems in the mammalian superior colliculus. Trends in neurosciences, 12(4), 137‒147. https://doi.org/10.1016/0166-2236(89)90052-0

[11] Goetz, L., Piallat, B., Bhattacharjee, M., Mathieu, H., David, O., & Chabardès, S. (2016). On the role of the pedunculopontine nucleus and mesencephalic reticular formation in locomotion in nonhuman primates. Journal of Neuroscience, 36, 4917‒4929. https://doi.org/10.1523/JNEUROSCI.2514-15.2016

[12] Milner, K. L., & Mogenson, G. J. (1988). Electrical and chemical activation of the mesencephalic and subthalamic locomotor regions in freely moving rats. Brain research, 452(1-2), 273‒285. https://doi.org/10.1016/0006-8993(88)90031-5

[13] Roseberry, T. K., Lee, A. M., Lalive, A. L., Wilbrecht, L., Bonci, A., & Kreitzer, A. C. (2016). Cell-type-specific control of brainstem locomotor circuits by basal ganglia. Cell, 164, 526‒537. https://doi.org/10.1016/j.cell.2015.12.037

[14] Josset, N., Roussel, M., Lemieux, M., Lafrance-Zoubga, D., Rastqar, A., & Bretzner, F. (2018). Distinct contributions of mesencephalic locomotor region nuclei to locomotor control in the freely behaving mouse. Current Biology, 28, 884–901.e3. https://doi.org/10.1016/j.cub.2018.02.007

[15] Kaur, J., Komi, S. A., Dmytriyeva, O., Houser, G. A., Bonfils, M. C. A., & Berg, R. W. (2025). Pedunculopontine-stimulation obstructs hippocampal theta rhythm and halts movement. Scientific reports, 15(1), 17903. https://doi.org/10.1038/s41598-025-01695-8

[16] Krauth, N., Sach, L. K., Sitzia, G., Clemmensen, C., & Kiehn, O. (2025). A hypothalamus-brainstem circuit governs the prioritization of safety over essential needs. Nature neuroscience, 28(7), 1473–1485. https://doi.org/10.1038/s41593-025-01975-6

[17] Lavoie, B., & Parent, A. (1994). Pedunculopontine nucleus in the squirrel monkey: Cholinergic and glutamatergic projections to the substantia nigra. Journal of Comparative Neurology, 344, 232–241. https://doi.org/10.1002/cne.903440205

[18] Zhang, S., Mena-Segovia, J., & Gut, N. K. (2024). Inhibitory Pedunculopontine Neurons Gate Dopamine-Mediated Motor Actions of Unsigned Valence. Current neuropharmacology, 22(9), 1540–1550. https://doi.org/10.2174/1570159X21666230911103520

[19] Futami, T., Takakusaki, K., & Kitai, S. T. (1995). Glutamatergic and cholinergic inputs from the pedunculopontine tegmental nucleus to dopamine neurons in the substantia nigra pars compacta. Neuroscience research, 21(4), 331–342.

https://doi.org/10.1016/0168-0102(94)00869-h

[20] Huerta-Ocampo, I., Dautan, D., Gut, N. K., Khan, B., & Mena-Segovia, J. (2021). Whole-brain mapping of monosynaptic inputs to midbrain cholinergic neurons. Scientific reports, 11(1), 9055. https://doi.org/10.1038/s41598-021-88374-6

[21] Floresco, S. B., West, A. R., Ash, B., Moore, H., & Grace, A. A. (2003). Afferent modulation of dopamine neuron firing differentially regulates tonic and phasic dopamine transmission. Nature Neuroscience, 6, 968–973.

https://doi.org/10.1038/nn1103

[22] Beier, K. T., Steinberg, E. E., DeLoach, K. E., Xie, S., Miyamichi, K., Schwarz, L., Gao, X. J., Kremer, E. J., Malenka, R. C., & Luo, L. (2015). Circuit Architecture of VTA Dopamine Neurons Revealed by Systematic Input-Output Mapping. Cell, 162(3), 622–634. https://doi.org/10.1016/j.cell.2015.07.015

[23] Stuber, G. D., Britt, J. P., & Bonci, A. (2012). Optogenetic modulation of neural circuits that underlie reward seeking. Biological Psychiatry, 71, 1061−1067.

https://doi.org/10.1016/j.biopsych.2011.11.010

[24] Stefanik, M. T., & Kalivas, P. W. (2013). Optogenetic dissection of basolateral amygdala projections during cue-induced reinstatement of cocaine seeking. Frontiers in Behavioral Neuroscience, 7, 213. https://doi.org/10.3389/fnbeh.2013.00213

[25] Cao, Z. F., Burdakov, D., & Sarnyai, Z. (2011). Optogenetics: potentials for addiction research. Addiction biology, 16(4), 519–531. https://doi.org/10.1111/j.1369-1600.2011.00386.x

[26] Sandler, M., Howard, A., Zhu, M., Zhmoginov, A., & Chen, L.-C. (2018). MobileNetV2: Inverted residuals and linear bottlenecks. Proceedings of the IEEE Conference on Computer Vision and Pattern Recognition, 4510–4520. https://doi.org/10.1109/CVPR.2018.00474

[27] Nguyen H. Fast object detection framework based on MobileNetV2 architecture and enhanced feature pyramid. J Theor Appl Inf Technol 2020;98:812–23. https://api.semanticscholar.org/CorpusID:219619148

[28] Jongboom, J. FOMO: Real-Time Object Detection on Microcontrollers. Available online: https://docs.edgeimpulse.com/docs/fomo-object-detection-for-constrained-devices

[29] Li, G., Komi, S., Sørensen, J. F., & Berg, R. W. (2025). A real-time vision-based adaptive follow treadmill for animal gait analysis. Sensors, 25(14), 4289. https://doi.org/10.3390/s25144289

[30] OpenMV. A Python Powered, Extensible Machine Vision Camera. Available online: https://openmv.io

[31] Iqbal H. PlotNeuralNet: Latex Code for Drawing Neural Networks. GitHub Repository, 2019. Available online: https://github.com/HarisIqbal88/PlotNeuralNet.

[32] Okada, K., Toyama, K., Inoue, Y., Isa, T., & Kobayashi, Y. (2009). Different pedunculopontine tegmental neurons signal predicted and actual task rewards. Journal of Neuroscience, 29, 4858–4870. https://doi.org/10.1523/JNEUROSCI.4415-08.2009

[33] Pan, W. X., & Hyland, B. I. (2005). Pedunculopontine tegmental nucleus controls conditioned responses of midbrain dopamine neurons in behaving rats. Journal of

Neuroscience, 25, 4725–4732. https://doi.org/10.1523/JNEUROSCI.0277-05.2005

[34] Chettih, S. N., & Harvey, C. D. (2019). Single-neuron perturbations reveal feature-specific competition in V1. Nature, 567(7748), 334–340. https://doi.org/10.1038/s41586-019-0997-6

[35] von Ziegler, L., Sturman, O., & Bohacek, J. (2021). Big behavior: challenges and opportunities in a new era of deep behavior profiling. Neuropsychopharmacology : official publication of the American College of Neuropsychopharmacology, 46(1), 33–44. https://doi.org/10.1038/s41386-020-0751-7

[36] Brown R. E. (2024). Measuring the replicability of our own research. Journal of neuroscience methods, 406, 110111. https://doi.org/10.1016/j.jneumeth.2024.110111

[37] Zhang, G.-W., Shen, L., Li, Z., Tao, H. W., & Zhang, L. I. (2019). Track-Control: An automatic video-based real-time closed-loop behavioral control toolbox. bioRxiv [Preprint], 2019.12.11.873372. https://doi.org/10.1101/2019.12.11.873372

[38] Crabbe, J. C., Wahlsten, D., & Dudek, B. C. (1999). Genetics of mouse behavior: Interactions with laboratory environment. Science, 284, 1670–1672. https://doi.org/10.1126/science.284.5420.1670

[39] Mandillo, S., Tucci, V., Hölter, S. M., Meziane, H., Banchaabouchi, M. A., Kallnik, M., ... Wurst, W. (2008). Reliability, robustness, and reproducibility in mouse behavioral phenotyping: A cross-laboratory study. Physiological Genomics, 34, 243–255. https://doi.org/10.1152/physiolgenomics.90207.2008

[40] Thevathasan, W., Coyne, T. J., Hyam, J. A., Kerr, G., Jenkinson, N., Aziz, T. Z., ... Silburn, P. A. (2011). Pedunculopontine nucleus stimulation improves gait freezing in Parkinson disease. Neurosurgery, 69, 1248–1253. https://doi.org/10.1227/NEU.0b013e31822b6f71

[41] Tykocki T, Mandat T, Nauman P. Pedunculopontine nucleus deep brain stimulation in Parkinson's disease. Arch Med Sci 2011;7:555–64.Tykocki, T., Mandat, T., & Nauman, P. (2011). Pedunculopontine nucleus deep brain stimulation in Parkinson's disease. Archives of medical science : AMS, 7(4), 555–564. https://doi.org/10.5114/aoms.2011.24119

[42] Yu, K., Ren, Z., Guo, S., Li, J., & Li, Y. (2020). Effects of pedunculopontine nucleus deep brain stimulation on gait disorders in Parkinsonʹs disease: A meta-analysis. Clinical Neurology and Neurosurgery, 198, 106108. https://doi.org/10.1016/j.clineuro.2020.106108

[43] Thevathasan, W., Debu, B., Aziz, T., Bloem, B. R., Blahak, C., Butson, C., ... Moro, E. (2018). Pedunculopontine nucleus deep brain stimulation in Parkinsonʹs disease: A clinical review. Movement Disorders, 33, 10–20. https://doi.org/10.1002/mds.27098

[44] Gut, N. K., & Winn, P. (2016). The pedunculopontine tegmental nucleus-A functional hypothesis from the comparative literature. Movement disorders : official journal of the Movement Disorder Society, 31(5), 615–624. https://doi.org/10.1002/mds.26556

[45] Lau, B., Welter, M. L., Belaid, H., Fernandez, V. S., Bardinet, E., Grabli, D., ...

Karachi, C. (2015). The integrative role of the pedunculopontine nucleus in human gait. Brain, 138, 1284–1296. https://doi.org/10.1093/brain/awv047

[46] Takakusaki, K. (2017). Functional neuroanatomy for posture and gait control. Journal of Movement Disorders, 10, 1 – 17. https://doi.org/10.14802/jmd.16062

[47] Okada, K., & Kobayashi, Y. (2013). Reward prediction-related increases and decreases in tonic neuronal activity of the pedunculopontine tegmental nucleus. Frontiers in Integrative Neuroscience, 7, 36. https://doi.org/10.3389/fnint.2013.00036

[48] Zhang, S., Mena-Segovia, J., & Gut, N. K. (2024). Inhibitory Pedunculopontine Neurons Gate Dopamine-Mediated Motor Actions of Unsigned Valence. Current neuropharmacology, 22(9), 1540–1550. https://doi.org/10.2174/1570159X21666230911103520

# Supplementary material

Note: Rat111 and Rat183 correspond to "rat1" and "rat2" in the original manuscript, while Rat187 and Rat189 correspond to "rat3" and "rat4," respectively.

To evaluate the model's capacity to recognize biologically meaningful events, we focused on the classification performance of two critical classes which associated with optogenetic stimulation which was referred to as STI (the STI class was labeled as "RIR" in the training logs for Rat111 and Rat183, and as "rat" for Rat187 and Rat189, respectively). The "RIR" (region-of-interest entry events) was evaluated for Rat111 and Rat183, while the "rat" class (presence of rat in one arm of the Y maze) was analyzed for Rat187 and Rat189. As shown in Figure 1, a comparison of INT8 quantized models showed that both pairs demonstrated consistently high precision and recall values above 0.95, suggesting reliable detection. For the "RIR" class, models trained on Rat111 and Rat183 logs exhibited excellent performance, with F1 scores of 0.979 and 0.990, respectively. Notably, Rat183 achieved perfect recall (1.0), indicating the model detected all relevant instances. Precision remained high in both models (>0.98), confirming low false positive rates. In the "rat" class comparison, models trained on Rat187 and Rat189 demonstrated equally strong outcomes, both achieving F1 scores ≥0.979. Impressively, the Rat189 model attained perfect precision and recall, suggesting highly robust generalization. These metrics support the reliability of using quantized, low-resource models for triggering real-time stimulation tasks in embedded behavioral systems. The detailed performance of these models are summarized in Table 1. Overall, the metrics indicate the model's reliable detection across individual animals and its potential for real-time behavioral monitoring in closed-loop experimental setups. The background classes were excluded from this analysis to focus strictly on behaviorally relevant events.

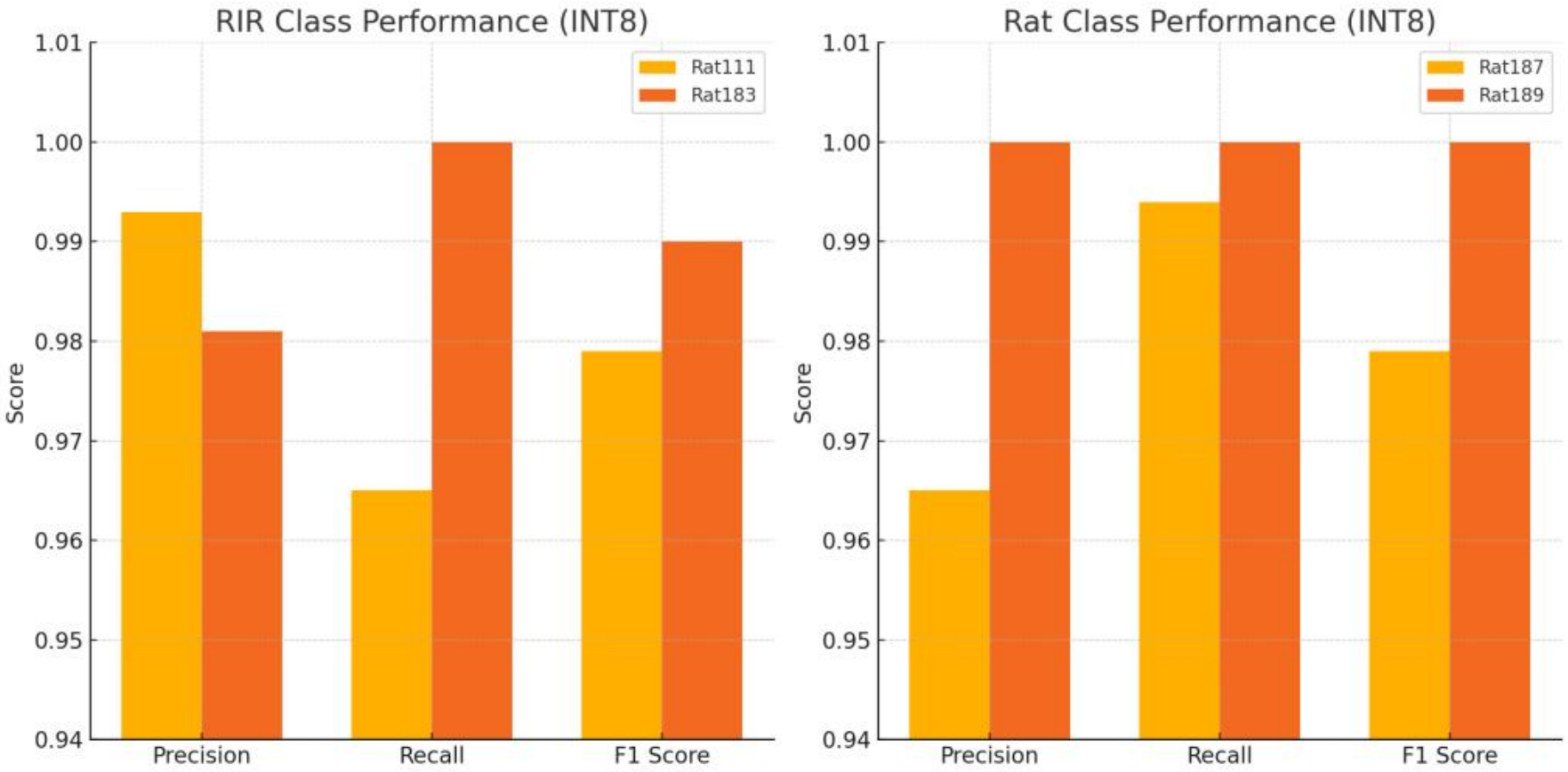

Figure 1. Performance comparison of stimulation-relevant classes (INT8 Quantized Models). Left: RIR class metrics for Rat111 vs. Rat183. Right: "Rat" class metrics for Rat187 vs. Rat189. Metrics include Precision, Recall, and F1 Score, highlighting detection performance across animals for STI relevant events.

TABLE I. Performance Summary Table for the "STI" Class (INT8)

| Model | Rat111 (RIR) | Rat183 (RIR) | Rat187 (rat) | Rat189 (rat) |
|---|---|---|---|---|
| **Precision** | 0.993 | 0.981 | 0.965 | 1.0 |
| **Recall** | 0.965 | 1.0 | 0.994 | 1.0 |
| **F1 Score** | 0.979 | 0.990 | 0.979 | 1.0 |

The training metrics summary for the "STI" class across all four rats using the INT8 quantized models trained with FOMO MobileNetV2 network.

Table 1 include Precision, Recall, and F1 Score metrics, providing insight into the consistency and accuracy of each model across subjects.

TABLE Ⅱ. Laser stimulation parameters

| Frequency | Duty cycle | Duration | Pause time | Repetitions | Trigger | Trigger delay | Trigger mode |
|---|---|---|---|---|---|---|---|
| 20Hz | 50% | 3000ms | 1ms | 1 | √ | 0ms | 2 |

The laser stimulus is controlled by specifying its pulse-train frequency (Hz), duty cycle (%), individual pulse duration (ms), inter-pulse pause time (ms), number of repetitions per bout (“1” means the train is delivered once), whether external triggering is used, the trigger delay (ms), and the trigger mode (“2” indicates a repeated pulse train, as opposed to single-shot mode).